 \documentclass[12pt,preprint]{aastex}

\shorttitle{Massive Pulsars and Ultraluminous X-ray Sources}
\shortauthors{Guo et al.}

\begin{document}


\title{Massive Pulsars and Ultraluminous X-ray Sources}


\author{Y. J. Guo\altaffilmark{1}, H. Tong\altaffilmark{2}, R. X. Xu\altaffilmark{1,3}
}
\affil{$^1$School of Physics and State Key Laboratory of Nuclear Physics and Technology, Peking University,
Beijing 100871, China; {\tt guoyj10@pku.edu.cn}}

\affil{$^2$Xinjiang Astronomical Observatory, Chinese Academy of Sciences, Urumqi 830011, China}

\affil{$^3$Kavli Institute for Astronomy and Astrophysics, Peking University, Beijing 100871, China}




\begin{abstract}

The detection of 1.37$\, $s pulsations from NuSTAR J095551+6940.8 implies the existence of an accreting pulsar, which challenges the conventional understanding of ultraluminous X-ray source.
%
%
%
This kind of sources are proposed to be massive pulsars in this paper.
Considering the general relativistic effect, stronger gravity of massive pulsars could lead to a larger maximum luminosity, scaled as the Eddington luminosity.
The pseudo-Newtonian potential is employed to simulate the gravitational field in general relativity, and the Eddington luminosity is calculated for self-bound stars (quark star and quark-cluster star) and for the Tolman IV solution.
It is found that, for a massive pulsar with radius close to the Schwarszchild radius, the Eddington luminosity could be as high as $2\times10^{39} \, {\rm erg\, s}^{-1}$.
It is able to account for the X-ray luminosity of NuSTAR J095551+6940.8 with reasonable beaming factor.
It is also suggested that massive pulsar-like compact stars could form via this super-Eddington phase of ultraluminous X-ray source.

\end{abstract}


\keywords{accretion --- dense matter --- pulsars: individual (NuSTAR J095551+6940.8) --- stars: magnetar --- stars: neutron }



\section{Introduction}

%
%

Ultraluminous X-ray sources (ULXs) have X-ray luminosity larger than $10^{39} \, {\rm erg\, s}^{-1}$, which exceeds the Eddington limit of stellar-mass objects~\citep{Feng2011}.
They are assumed to be accreting black holes, either intermediate-mass black holes accreting at sub-Eddington rate, or stellar-mass black holes with super-Eddington accretion.
Recently, an ULX in M82 was identified as a pulsar (NuSTAR J095551+6940.8) by ~\citet{Bachetti2014}, with an average period of 1.37$\, $s and a 2.5-day sinusoidal modulation.
The object has period derivative of $-2\times 10^{-10}\, $s and isotropic X-ray luminosity around $10^{40} \, {\rm erg\, s}^{-1}$.
Considering possible anisotropy in accretion of pulsar, one may think that the actual luminosity would be several times of  $10^{39} \, {\rm erg\, s}^{-1}$, which is still well above the conventional Eddington luminosity $\sim 2 \times 10^{38} \, {\rm erg\, s}^{-1}$ for an accreting pulsar with $M=1.4 M_\odot$ ($M_\odot$ is the solar mass).

The critical luminosity of an accreting compact object is determined by the balance of gravity and radiation pressure.
Both suppressing radiation pressure or increasing gravity would result in a larger critical luminosity.
To explain the super-Eddington phenemona, on one hand, NuSTAR J095551+6940.8 is suggested to be an accreting magnetar, which invokes super-strong multiple magnetic field to suppress the radiation pressure~\citep{Eksi2014, Lyutikov2014, TH2015}.
On the other hand, increasing the gravity is also feasible, that is focused in this paper.
%
%
The gravity in general relativity is effectively stronger than the Newtonian one in the limit of strong field, which could lead to a much higher Eddington luminosity in principle.
This general relativistic effect would be significant only for pulsar with large mass and small radius so that the real radius is approaching the Schwarzchild one.
Generally, the radius of self-bound star is smaller than that of gravity-bound star, so self-bound stars would have larger Eddington luminosity than gravity-bound star with the same mass.
Additionally, the critical luminosity of quark-cluster star could even be higher due to high maximum mass \citep[e.g., see Fig. 6 in][]{Li2015}.

The modification of Eddington luminosity due to general relativistic effect is calculated in \S2, for two classes of solutions to Einstein's field equation.
One is numerical solution based on equation of state (EoS) for quark star and quark-cluster star, and the other is the analytical Tolman IV solution~\citep{Tolman39}.
The results show that massive pulsars could have Eddington luminosity as large as $2\times10^{39} \, {\rm erg\, s}^{-1}$, which might be able to explain the X-ray luminosity of NuSTAR J095551+6940.8.

High luminosity implies high accreting rate, and the central object is more probable to have large mass during its evolution.
This not only explains the ultraluminous X-ray pulsar, but also is consistent with the discovery of 2$M_\odot$ pulsar~\citep{Demorest2010,Antoniadis2013}.
The timescale for a $1.4M_\odot$ pulsar to reach maximum mass by accretion is also estimated, which has the same scale as that of binary evolution.
The spin-down rate of NuSTAR J095551+6940.8 might be similar to an accreting pulsar with normal magnetic field of $\sim 10^{12}$ G~\citep{TH2015}.
Thus NuSTAR J095551+6940.8 could be an accreting massive quark star or quark-cluster star, without the need for strong magnetic field in the magnetar models.
If future observations could find evidence for whether strong magnetic field exists or not, it will be possible to distinguish accreting self-bound star and accreting magnetar.
%
%
%

%
The modified Eddington luminosity in strong gravity is calculated in \S2.
In \S3, the spin behavior and the timescale of accretion are discussed.
Conclusions are made in \S4.

\section{Eddington luminosity with pseudo-Newtonian potential}

\subsection{Modified Eddington luminosity}

A pseudo-Newtonian potential $\Psi=-GM/(r-R_{\rm s})$ (the Schwarzschild radius $R_{\rm s}=2GM/c^2$) is used to simulate the gravitational field of the Schwarzschild metric, as first proposed by \citet{Paczynski1980}.
The potential together with the Newtonian equation of motion could have result which is in good agreement with that of  general relativity~\citep{Lu1985}, and largely simplify the calculation.
So it has been widely used when dealing with black hole accretion disks.
The Eddington luminosity is derived by balancing the gravity $f_{\rm g}$ and radiation pressure $f_{\rm r}$.
In the case of fully ionized hydrogen plasma, $f_{\rm g}=-m_{\rm p}\nabla \Psi = -GMm_{\rm p}/(r-R_{\rm s})^2$, and $f_{\rm r} = \rho \sigma$, where $m_{\rm p}$ is the proton mass, $\rho$ is the energy density of radiation field, and $\sigma$ is the Thomson cross section $\sigma_{\rm T}$ with non-relativistic electrons and weak magnetic field.
From $f_{\rm g} + f_{\rm r}=0$, we have $\rho= GMm_{\rm p}/(r-R_{\rm s})^2 \sigma_{\rm T}$, and the maximum luminosity (i.e., modified Eddington luminosity) emitted from the star's surface, for the case of isostropic accretion, is then
\begin{equation}
L_{\rm Edd,\, s}^\prime=\rho c \times 4\pi r^2
= \frac{4 \pi m_{\rm p} cGM}{\sigma_{\rm T}} \times \frac{r^2} {(r-R_{\rm s})^2}
=L_{\rm Edd}\times (1-R_{\rm s}/r)^{-2}.
\end{equation}
As a result of gravitational redshift, however, the luminosity observed at infinity is,
\begin{equation}
L_{\rm Edd}^\prime=L_{\rm Edd,\, s}^\prime \times (1-R_{\rm s}/r) = L_{\rm Edd}\times (1-R_{\rm s}/r)^{-1}.
\end{equation}

Compared with the conventional Eddington luminosity $L_{\rm Edd}=4 \pi m_{\rm p} cGM / \sigma_{\rm T}$, there is a correction factor $(1-R_{\rm s}/r)^{-1}$ in the modified Eddington luminosity $L_{\rm Edd}^\prime$.
The correction factor $(1-R_{\rm s}/r)^{-1}$ increases rapidly as $r/R_{\rm s}$ approaches 1.
When $r/R_{\rm s}<9/8$, the central density of star would be infinite~\citep{Buchdahl1959}, so an upper limit is set to the correction factor, $(1-R_{\rm s}/r)^{-1}<9$.
The correction factor is unimportant at large $r$; it has a value of 2 when $r/R_{\rm s}=2$, and 1.25 when  $r/R_{\rm s}=5$.
It is then worth noting that, if a star is extremely compact ($r$ is small), its $L_{\rm Edd}^\prime$ could be one order of magnitude larger than $L_{\rm Edd}$.

Moreover, maximum luminosity approaching or even higher than $L_{\rm Edd}^\prime$ can only be possible in case of anisotropic accretion.
In case of isotropic accretion, the general luminosity cannot be larger than $L_{\rm Edd}$ with which the gravity and radiation can certainly balance at sufficiently large distance where Newtonian gravity works (note: $L_{\rm Edd}$ is independent of distance $r$, while $L_{\rm Edd}^\prime$ decreases as $r$ increases).
Nevertheless, if the spherical symmetry is broken, radiative photons could be opaque in the direction of accretion flow but transparent in other ways so that the balance of radiation pressure and gravity would be achieved around the star surface where the radiation pressure would be stronger than gravity at larger radius.
%
%
Inside the accreted fluid, the radiation intensity should decrease with increasing distance.
The radiation pressure could be weakened by the above factors, and balance of pressure and gravity at larger radius is then possible.
Therefore, the value of  $L_{\rm Edd}'$ at the radius $R$ of the central compact star is chosen to scale the star's maximum accretion rate.

The modification of Eddington luminosity due to general relativistic effect is discussed for mass-radius relations in two cases.
One is numerical result based on EoS for quark star and quark-cluster star, and the other is the analytical Tolman IV solution to Einstein's field equation~\citep{Tolman39}.

\subsection{Case I: quark-cluster star and quark star}

The EoS of dense matter at supra-nuclear density is still not well known because this key question is essentially of non-perturbative quantum chromo-dynamics in the regime of low-energy scale \citep[e.g.,][]{LP2004}).
Among different models for the inner structure of pulsar, a quark-cluster star could naturally have very stiff equation of state~\citep{LX2009,LX2009b,Lai2013,GLX2014}, which is consistent with the later discoveries of massive pulsars~\citep{Demorest2010,Antoniadis2013}.
Recently, an analysis of the type I X-ray burst observations may show that the compact star in 4U 1746-37 could have an extremely low mass and small radius, that could also be understood in the quark-cluster star model~\citep{Li2015}.

Various models for pulsar can be divided into two classes, self-bound and gravity-bound stars.
Gravity-bound star includes conventional neutron star and hybrid star, the radius of which usually decreases as the mass increases.
Self-bound star includes quark star~\citep{Ioth1970, Alcock1986, Haensel1986} and quark-cluster star~\citep{Xu2003}, where the radius generally increases with mass.
The general relativistic effect would be more important for quark star and quark-cluster star, since they usually have smaller radius than gravity-bound star of the same mass.
The modified Eddington luminosity is calculated for several EoS of self-bound star.

\begin{figure}
\epsscale{.70}
\plotone{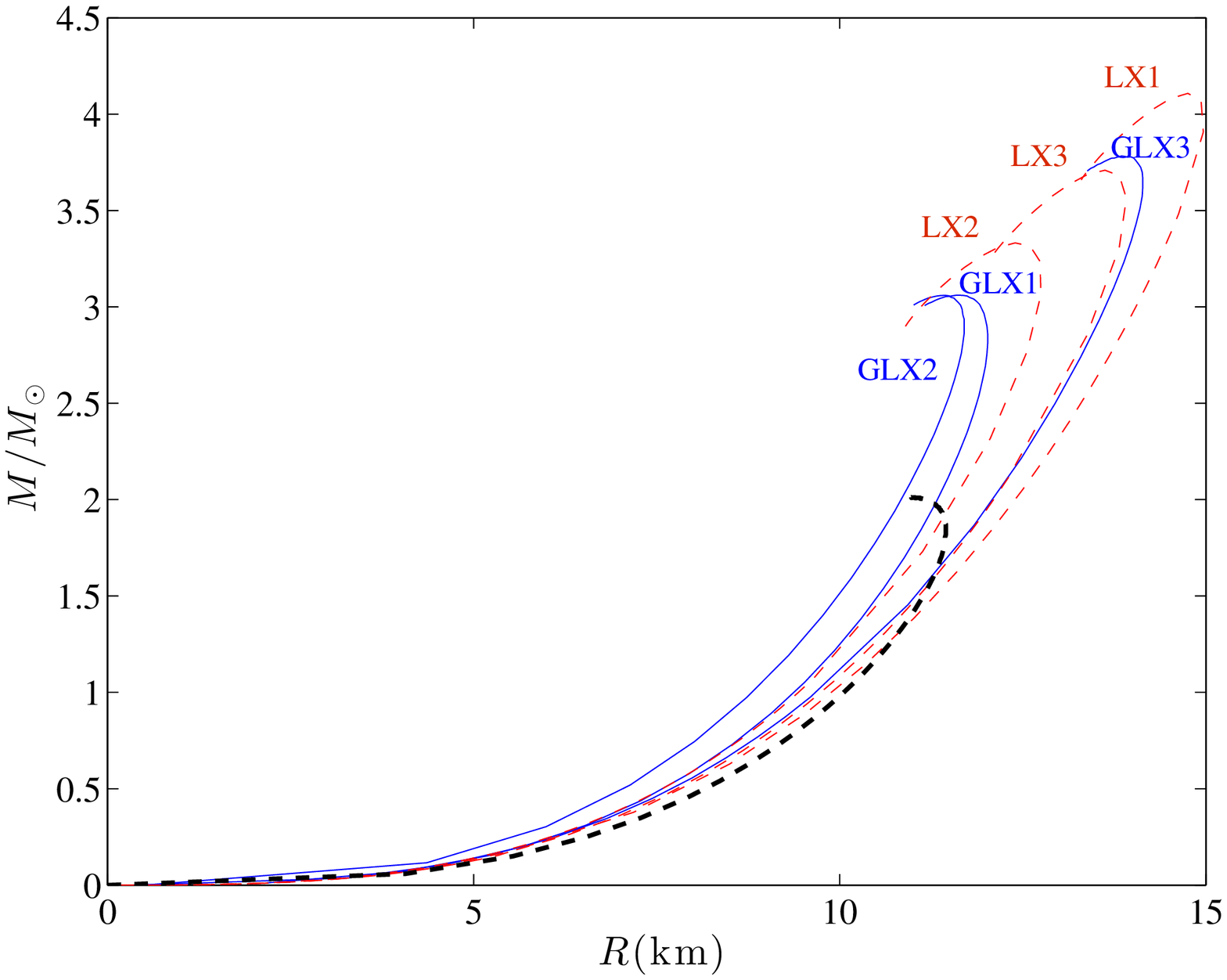}
\plotone{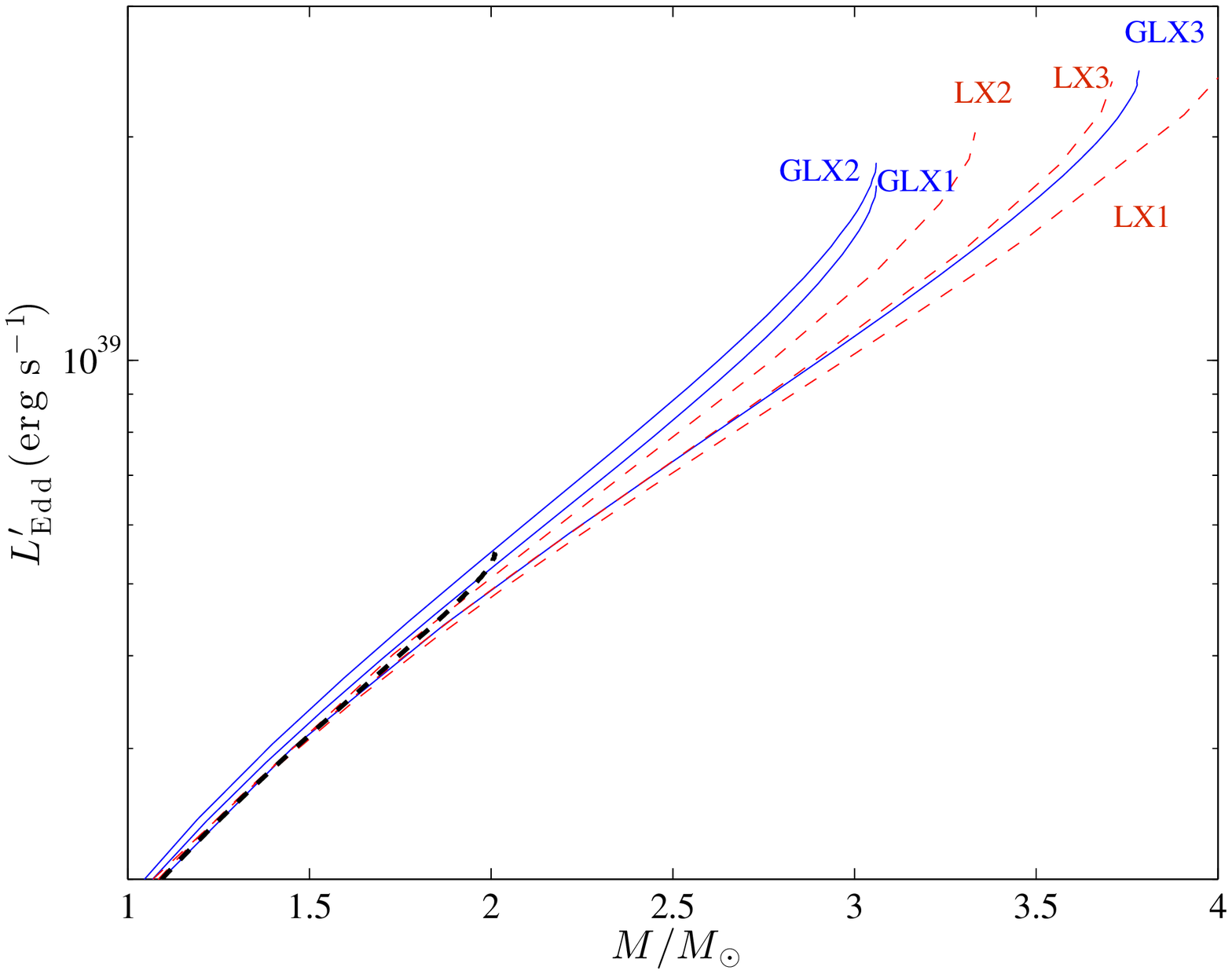}
\caption{Upper panel: the mass-radius curves for several EoS. Lower panel: the relation between the Eddington luminosity $L'_{\rm Edd}$ and mass $M$ for these EoS. Solid lines represent EoS from GLX~\citep{GLX2014}, and dashed lines represent EoS from LX~\citep{LX2009}. Different numbers indicate different parameters. The dot-dashed line is the result of MIT bag model (bag constant $B = 57\, {\rm MeV\, fm}^{-3}$).}\label{leos}
\end{figure}

The hydrostatic equilibrium equation in general relativity is the Tolman-Oppenheimer-Volkoff (TOV) equation~\citep{TOV1939},
\begin{equation}
  \frac{dP}{dr}=-\frac{Gm(r)\rho}{r^2}\frac{(1+\frac{P}{\rho c^2})(1+\frac{4\pi r^3P}{m(r)c^2})}{1-\frac{2Gm(r)}{rc^2}},
\end{equation}
where $m(r)=\int_0^r \rho \cdot 4\pi r'^2 dr'$.
Combine the TOV equation with the EoS, the mass and radius of the star can be calculated by numerical integration.
For quark star, the simple MIT bag model is chosen.
For quark-cluster star, Lennard-Jones potential~\citep{LX2009} and a corresponding-state approach~\citep{GLX2014} have been used to derive the EoS.
The modified Eddington luminosity $L'_{\rm Edd}$ is calculated for the mass-radius relations given by the above EoS, and the results are plotted against mass in Figure~\ref{leos}.
When the mass is small, $L'_{\rm Edd}$ is similar for different EoS.
The MIT bag constant is chosen to be $57\, {\rm MeV\, fm}^{-3}$ so that we have has a maximum mass of $\sim 2M_\odot$, and the corresponding critical luminosity is $\sim 5\times 10^{38} \, {\rm erg\, s}^{-1}$.
Quark-cluster stars have maximum mass in the range of $3-4M_\odot$, and $L'_{\rm Edd}$ could be as high as $2\times10^{39} \, {\rm erg\, s}^{-1}$.

\subsection{Case II: Tolman IV solution}

Besides the numerical solutions employing physically motivated EoS, analytical solutions to Einstein's field equation can also be obtained with appropriate conditions on the metric.
This mathematically motivated method was first suggested by Tolman, and a number of new solutions were obtained, among which Tolman IV solution is suitable to describe self-bound star~\citep{Tolman39}.
Based on the mass and radius given by the Tolman IV solution, $L'_{\rm Edd}$ is also calculated, for more general results independent of detailed EoS.

For static and spherical distribution of matter, the line element can be written in the simple form,
\begin{equation}
ds^2 = -e^\lambda dr^2 - r^2 d\theta^2 - r^2 \sin^2\theta d\phi^2 + e^\nu dt^2 .
\end{equation}
Tolman IV solution stems from the assumption,
\begin{equation}
e^\nu \nu^\prime / 2r = {\rm const} ,
\end{equation}
and the result is determined by two independent parameters A and C.
The central density $\rho_{\rm c}$, central pressure $p_{\rm c}$, radius $R$, surface density $\rho_{\rm b}$ and mass $M$ are
\begin{eqnarray}
8\pi \rho_{\rm c} & = & \frac{3}{A^2} + \frac{3}{C^2} , \ \ {\rm and} \ \   8\pi p_{\rm c} = \frac{1}{A^2} - \frac{1}{C^2}, \\
R & = & \frac{C}{3^{1/2}} (1- A^2/C^2) ^ {1/2} , \label{R} \\
\rho_{\rm b} & = & \rho_{\rm c} - 5 p_{\rm c} + 8p_{\rm c}^2/(\rho_{\rm c}+p_{\rm c}) ,  \\
M & = & \frac{R}{2} \left \{ 1-\frac{(1-R^2/C^2)(1+R^2/A^2)}{1+2R^2/A^2} \right \} , \label{M}
\end{eqnarray}
respectively (in this subsection, the $G=c=1$ units are used).
From equation~(\ref{R}) and equation~(\ref{M}),
\begin{eqnarray}
2M/R & = & 1-\frac{(1-R^2/C^2)(1+R^2/A^2)}{1+2R^2/A^2} \nonumber \\
          & = & \frac{2}{3} (1 - A^2/C^2) < \frac{2}{3} .
\end{eqnarray}
So there is a lower limit $R/2M > 1.5$, and the correction factor $(1-R_s/R)^{-2}=(1-2M/R)^{-2}$ could not be larger than 3.

\begin{figure}
\epsscale{.70}
\plotone{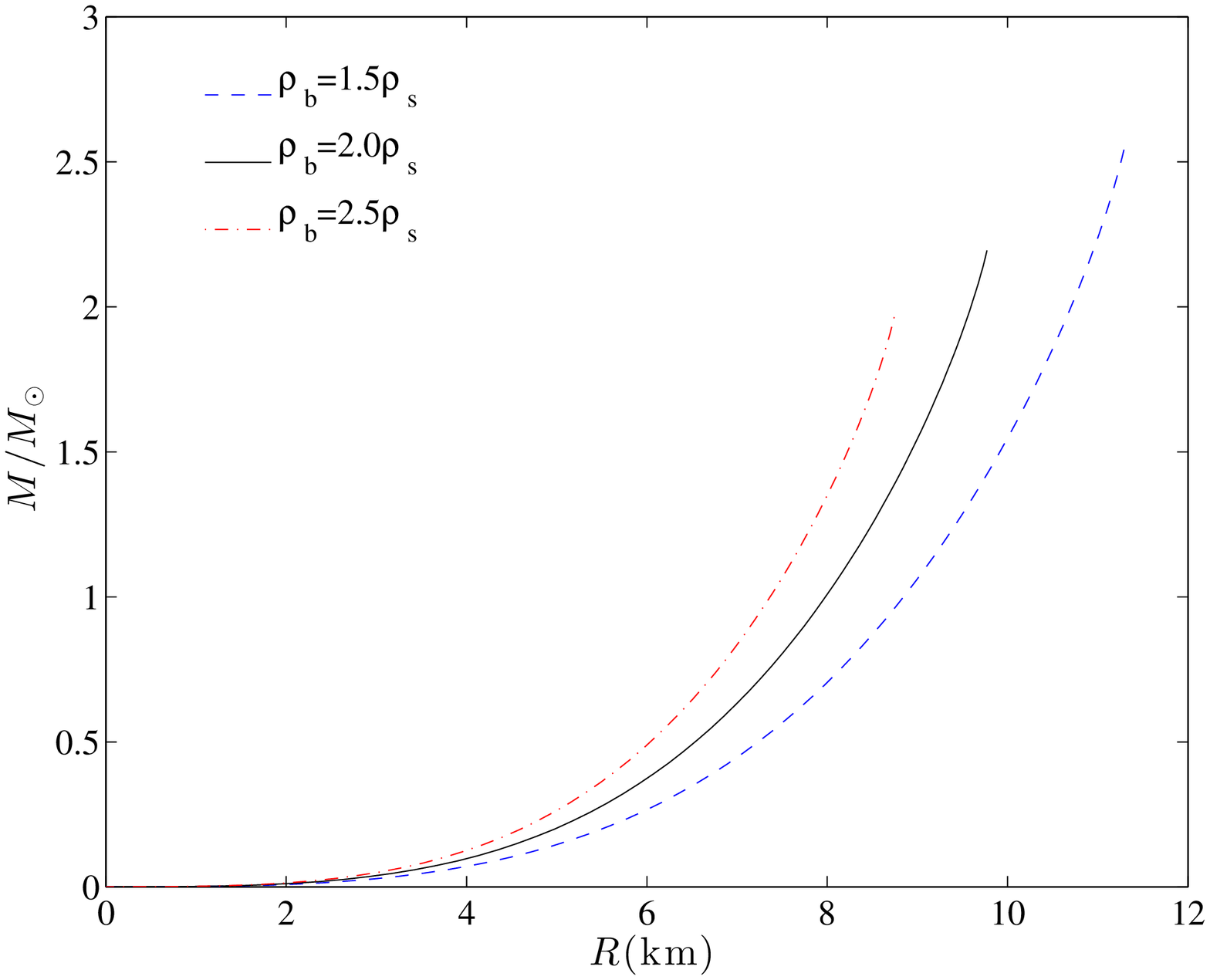}
\plotone{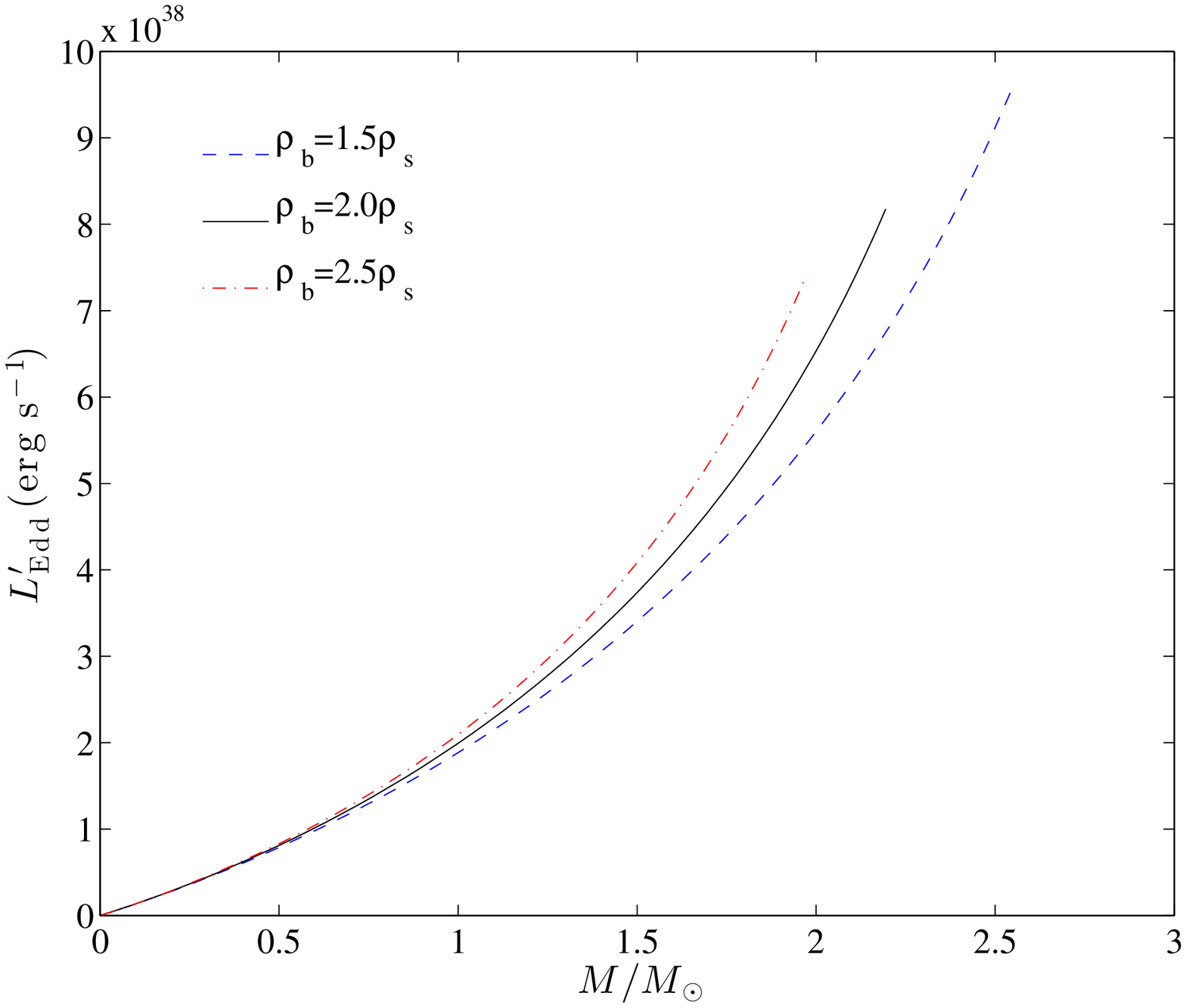}
\caption{Upper panel: the mass $M$ and radius $R$ given by different values of parameter $A$ in Tolman IV solution. Lower panel: the relation between the modified Eddington luminosity and mass for Tolman IV solution. Different lines represent three cases of surface density $\rho_{\rm b}$ --- $1.5\rho_{\rm s}$ (dashed line), $2\rho_{\rm s}$ (solid line) and $2.5 \rho_{\rm s}$ (dot-dashed line).}\label{tov}
\end{figure}

Since there are two independent parameters, the value of surface density is fixed to reduce one degree of freedom, and thus all the properties can be regarded as functions of the single parameter A.
For self-bound star with finite surface density, three cases $\rho_{\rm b}=1.5\rho_{\rm s}, 2\rho_{\rm s}, 2.5 \rho_{\rm s}$ are considered, where $\rho_{\rm s}$ is the nuclear saturation density.
The resulting pairs of $M$ and $R$ are plotted in Figure~\ref{tov}, and the modified Eddington luminosity for different mass is also shown.
For Tolman IV solution, $L'_{\rm Edd}$ is smaller than that in the case of quark-cluster star.
If considering the beaming effect, it might as well explain the X-ray luminosity of NuSTAR J095551+6940.8.

\section{Discussions}

%
%
NuSTAR J095551+6940.8 has a spin period of $P=1.37\, {\rm s}$ and a period derivative of ${\dot P} \approx -2 \times 10^{-10}$.
The pulsar is spinning up and gaining angular momentum from the accretion flow.
According to the conservation of angular momentum, the strength of dipole magnetic filed $B_{\rm p}$ can be estimated.
%
%
%
%
%
%
%
%
%
%
%
For a pulsar with mass of $2M_\odot$ and beaming factor of 0.2, the value of $B_{\rm p}$ is $1.6 \times 10^{12} R_6 L_{40}^{-3}\, $G, where $R_6=R/10^6 \, {\rm cm}$ and $L_{40}$ is the X-ray luminosity in units of $10^{40}\, {\rm erg\, s}^{-1}$~\citep{TH2015}.
So the spin evolution of NuSTAR J095551+6940.8 might be similar to an accreting pulsar with normal magnetic field of $\sim 10^{12}$ G, and there might be no need for strong magnetic field being necessary in the magnetar models.

The birth rate of accreting quark star or quark-cluster star and accreting magnetar would be different, which might be studied in the future work.
This could answer the question, what proportion does pulsar system take up in the ULX population? What is the nature of ULX pulsar, accreting self-bound star or accreting magnetar?
The problem is also related to the nature of anomalous X-ray pulsar/soft Gamma-ray repeater: are they magnetar or ``quark star/quark-cluster star + fallback disk system"~\citep{TX2011}?

\subsection{Accretion process and formation of massive pulsars}

\begin{figure}
\epsscale{.70}
\plotone{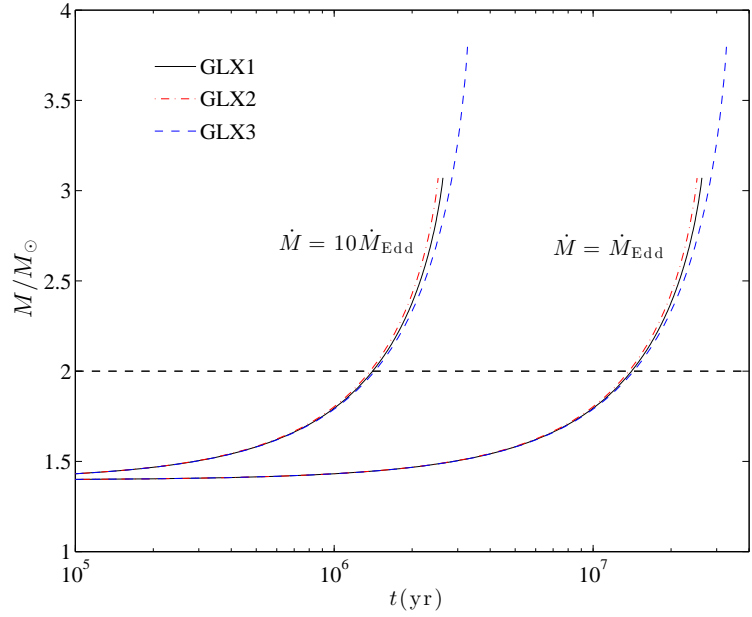}

\caption{The growth of mass with time at accretion rate of $\dot{M}_{\rm Edd}$ and $10\dot{M}_{\rm Edd}$, from 1.4 $M_\odot$ to the  maximum mass. Different lines are the results of different EoS~\citep{GLX2014}. The horizontal line detonates the position of 2$M_\odot$.}\label{mass_grow}
\end{figure}

From calculations in \S 2, only massive pulsars above $2.5 M_\odot$ could have critical luminosity larger than $\sim 10^{39} \, {\rm erg\, s}^{-1}$.
From the perspective of evolution, the pulsar is also more probable to have large mass since high luminosity indicates high accretion rate.
When born in supernova explosion, the pulsar may probably have a mass of 1.4$M_\odot$ only.
Accretion from companion star may lead to the formation of massive pulsars with mass of $2M_\odot$ or even higher.
According to mass-radius relation in ~\citet{GLX2014}, the time scale is estimated as following for a $1.4M_\odot$ pulsar to reach the maximum mass with accretion rate of $\dot{M}_{\rm Edd}$, where $\dot{M}_{\rm Edd}$ is the critical accretion rate corresponding to $L'_{\rm Edd,\, s}$.
The accretion of pulsar might be anisotropic because of strong dipole magnetic field ($\sim 10^{12}\, {\rm G}$), and the critical luminosity would then be larger than the isotropic case due to geometric effect~\citep{Basko1976}.
Certainly, it could be possible that accretion toward an compact object occurs at super-Eddington luminosity in a binary system~\citep{Wiktorowicz}.
The situation of $10\dot{M}_{\rm Edd}$ accretion is also considered, and the results are shown in Figure~\ref{mass_grow}.
The timescale is $\sim 10^7$ years for $\dot{M}_{\rm Edd}$ and $\sim 10^6$ years for $10\dot{M}_{\rm Edd}$, which is consistent with that of binary evolution models~\citep{ShaoLi2015, Fragos2015}.

%
Consider a binary system composed of a high-mass star A ($\gtrsim 12 M_\odot$) and a low or intermediate-mass star B ($\sim 4 M_\odot$).
The high-mass star A first evolves and explodes to become a pulsar with mass of $1.4 M_\odot$, and the mass of companion B could be increased to $\sim 5 M_\odot$ through accretion
%
When star B expands and exceeds its Roche lobe, mass will be transferred to pulsar A via the Lagrangian point.
As the mass of pulsar approaches $2M_\odot$, the accretion rate would become larger and it behaves as an ULX pulsar.
Then in the common envelope stage, the orbit decays with the release of orbital gravitational energy and the envelope could be blown off.
Finally there leave a massive pulsar and a white dwarf of $\sim 0.5 M_\odot$.
The above scenario might explain both the discoveries of 2$M_\odot$ pulsar~\citep{Demorest2010,Antoniadis2013} and ULX pulsar~\citep{Bachetti2014}.

\section{Conclusion}

The identification of NuSTAR J095551+6940.8 as a pulsar challenges the understanding of accreting neutron star.
%
%
Considering the general relativistic effect, the Eddington luminosity of massive pulsars could be much larger than $10^{38} \, {\rm erg\, s}^{-1}$ in the frame of self-bound star.
For numerical solutions based on detailed EoS, the Eddington luminosity of quark-cluster star could be as high as $2 \times 10^{39} \, {\rm erg\, s}^{-1}$, and that of quark star can reach $5 \times 10^{38} \, {\rm erg\, s}^{-1}$.
For analytical solutions to Einstein's field equations, the Tolman IV solution gives Eddington luminosity up to $10^{39} \, {\rm erg\, s}^{-1}$.
The X-ray luminosity of NuSTAR J095551+6940.8 can be well explained by massive pulsar with critical accretion rate.

In addition, the spin behavior of NuSTAR J095551+6940.8 can be accounted for by normal magnetic filed.
The timescale for a $1.4M_\odot$ pulsar to reach maximum mass by accretion is consistent with that of binary evolution.
It is also proposed that as a binary system evolves, it might undergo the ULX phase in the accretion process, and finally a massive pulsar ($\ge 2M_\odot$) could be formed.

\acknowledgments

We would like to thank the pulsar group of PKU for useful discussions.
This work is supported by National Basic Research Program of China (973 program, 2012CB821801) and the National Natural Science
Foundation of China (Grant Nos. 11225314, 11133002, 11103020).




\clearpage

\end{document}